\documentclass[prl,aps,10pt,twocolumn,superscriptaddress,nofootinbib]{revtex4-1}

\usepackage{amssymb,amsmath}
\usepackage{xcolor}
\usepackage[utf8]{inputenc}

\usepackage{graphicx}

\usepackage[makeroom]{cancel}
\usepackage[caption=false]{subfig}
\usepackage[colorlinks=true,
            citecolor=green,
            linkcolor=red,
            filecolor=cyan,
            urlcolor=magenta,
            backref=false]{hyperref}

\usepackage{enumitem}

\newcommand{\dd}{\mbox{d}}

\begin{document}

\title{Note on black holes with kilometer-scale ultraviolet regulators}

\author{Jens Boos}
\email{jens.boos@kit.edu}
\affiliation{High Energy Theory Group, Department of Physics, William \& Mary, Williamsburg, VA 23187-8795, United States}
\affiliation{Institute for Theoretical Physics, Karlsruhe Institute of Technology, D-76128 Karlsruhe, Germany}

\author{Christopher D. Carone}
\email{cdcaro@wm.edu}
\affiliation{High Energy Theory Group, Department of Physics, William \& Mary, Williamsburg, VA 23187-8795, United States}

\date{May 30, 2024}

\begin{abstract}
Regular black hole metrics involve a universal, mass-independent regulator that can be up to $\mathcal{O}(700\,\text{km})$ while remaining consistent with terrestrial tests of Newtonian gravity and astrophysical tests of general relativistic orbits. However, for such large values of the regulator scale, the metric describes a compact, astrophysical-mass object with no horizon rather than a black hole.  We note that allowing the regulator to have a nontrivial mass dependence preserves the horizon, while allowing  large, percent-level effects in black hole observables.  By considering the deflection angle of light and the black hole shadow, we demonstrate this possibility explicitly.
\end{abstract}

\maketitle

\emph{Introduction.}---Astrophysical black holes are emerging as increasingly relevant testing grounds of gravitational physics \cite{Will:2014kxa,Berti:2015itd,Barack:2018yly,EventHorizonTelescope:2019dse,LIGOScientific:2020tif,Horowitz:2023xyl}. Under some assumptions on the energy and pressure of matter, general relativity predicts that black holes formed during the late stages of a collapsing massive star necessarily contain singularities \cite{Penrose:1964wq,Hawking:1970zqf,Hawking:1973uf}. These singularities show up as divergent gravitational tidal forces in the black hole interior, and prevent a complete description of physics in this regime \cite{Geroch:1968ut}. This is the black hole singularity problem, a major unresolved problem in gravitational physics. Within general relativity, these singularities are evident in the Schwarzschild metric: an exact, spherically symmetric vacuum solution of the Einstein equations, taking the form \cite{Schwarzschild:1916,Droste:1917}
\begin{align}
\begin{split}
\dd s^2 &= -F(r) \, \dd t^2 + \frac{\dd r^2}{F(r)} + r^2\dd\theta^2 + r^2\sin^2\theta \, \dd\varphi^2 \, , \\
F(r) &= 1 - \frac{2GM}{r} \, .
\end{split}
\end{align}
Here, $M$ is the black hole mass, $G$ is the gravitational constant, and we set the speed of light to unity ($c=1$). The event horizon is located at $r_\text{h} = 2GM$, and the singularity is situated at $r=0$, as can be seen from the Kretschmann invariant $R{}_{\mu\nu\rho\sigma}R{}^{\mu\nu\rho\sigma} = 48(GM)^2/r^6$. It is believed that quantum-gravitational effects change the form of the metric in the black hole interior at $r \ll r_\text{h}$, eventually leading to an avoidance of the singularity. In the recent years, a variety of regular black hole models have been proposed \cite{Bardeen:1968,Dymnikova:1992ux,Bonanno:2000ep,Hayward:2005gi,Simpson:2018tsi,Frolov:2021vbg}; see also the review \cite{Frolov:2016pav}. Typically, they feature a modification of an otherwise singular metric, parametrized by a length scale $\ell > 0$, that removes the singularity at $r=0$. These metrics are not solutions of a fundamental gravitational theory, but can help constrain the leading order observational consequences of regular black holes. A characteristic and popular model is the Hayward metric \cite{Hayward:2005gi},
\begin{align}
\label{eq:hayward}
F(r) &= 1 - \frac{2GM}{r} \frac{r^3}{r^3 + L^3} \, , \quad L^3 = 2GM\ell^2 \, .
\end{align}

\begin{figure*}[!htb]
\centering
\includegraphics[width=0.95\textwidth]{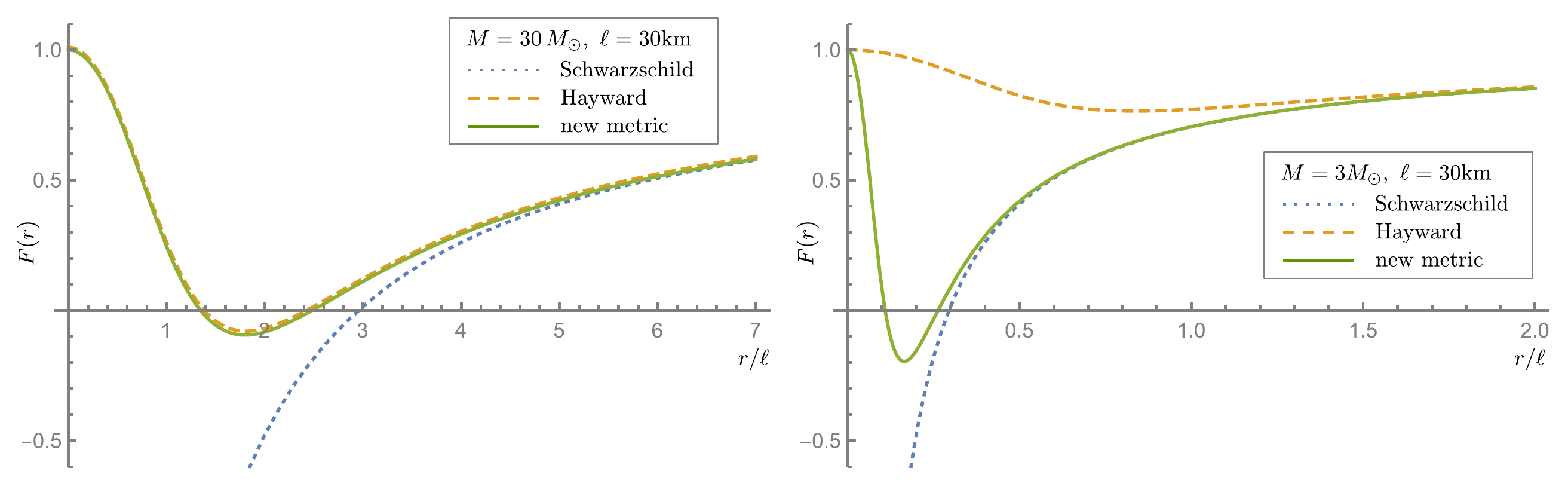}
\caption{A stark difference: The metric function $F(r)$ over the dimensionless distance $r/\ell$ for $\ell = 30\,\text{km}$ for two cases of $M$, assuming Eq.~\eqref{eq:example}. Left: $M = 30\,M_\odot$, the new metric is indistinguishable from the Hayward metric. Right: $M=3M_\odot$. The new metric features a horizon and deviates macroscopically from the Schwarzschild metric at the horizon by roughly 8\%; the Hayward metric does not describe a black hole, but a solar-mass scale horizonless object, so far unobserved in Nature.}
\label{fig:f}
\end{figure*}

For non-zero values of the parameter $L$, the metric in Eq.~(\ref{eq:hayward}) describes a smooth and finite gravitational field around $r=0$, while for $L=0$ one recovers the Schwarzschild solution. The parameter $L$ is the product of a universal length scale $\ell$ and a factor of $GM$ that removes the mass dependence of the function \eqref{eq:hayward} at small distances $r$. This universal behavior at small distances is known as the limiting curvature condition and is motivated by various high-energy properties of gravity~\cite{Markov:1982,Markov:1984,Polchinski:1989,Mukhanov:1991zn}. Hence we shall refer to $\ell$ (rather than $L$) as the fundamental, short-distance or ``ultraviolet'' (UV) regulator.

Let us now constrain both $L$ and $\ell$. E\"otv\"os-type experiments confirm the Newtonian inverse-square law of gravitation to distances of $\sim 50\,\mu\text{m}$ for test masses of $\mathcal{O}(1\,\text{kg})$. A measurement uncertainty of around $0.5\,\mu\text{m}$ \cite{Lee:2020zjt} bounds $L$ and $\ell$ by
\begin{align}
\label{eq:bound}
L \lesssim \mathcal{O}(10\,\mu\text{m}) \, , \qquad
\ell \lesssim \mathcal{O}(700\,\text{km}) \, .
\end{align}
The large separation between those two quantities stems from  $2GM/c^2 \sim 10^{-27}\,\text{m}$ for tabletop masses. Astrophysical orbits imply much weaker constraints. For a central, solar mass, at distances of $1\,\text{AU}$, the leading order deviation from the Schwarzschild metric is $(r_s/r)^2(\ell/r)^2 \sim 10^{-26}$, where $r_s = 2 G M /c^2$ and we set $M=M_\odot$ and $\ell = 700\,\text{km}$. For a supermassive black hole, such as Sagittarius A*, with $M \sim 4.3 \times 10^6 M_\odot$, a stellar object at its closest point of approach at $r = 120\,\text{AU}$ \cite{GRAVITY:2018ofz} would experience corrections of $(r_s/r)^2(\ell/r)^2 \sim 10^{-21}$, whereas $r_s/r \sim 10^{-3}.$

In the absence of a model for the origin of Eq.~\eqref{eq:hayward}, it is well motivated to treat $\ell$ as a free parameter and study the phenomenological consequences. 
The surprisingly weak bound in \eqref{eq:bound} implies that $\ell$ can be comparable to the horizon size of astrophysical black holes. When this is the case, regular black hole metrics can differ significantly in the near-horizon region from their Schwarzschild counterparts. The possibility of striking observational consequences, however, cannot be realized: the metric \eqref{eq:hayward} ceases to describe a black hole if $GM < (3\sqrt{3}/4) \ell$, preventing black holes with  masses below $\sim \ell/G$.\footnote{This remains true in dynamical black hole formation \cite{Frolov:2015bta}.} Importantly, this is not a fluke of the Hayward metric, but a common feature of \emph{all} known regular black hole metrics, usually referred to as a ``mass gap.'' See Table~\ref{table:1} for a summary of mass gaps for several well-known regular black hole models. To the best of our knowledge, no known regular black hole model can simultaneously accommodate a large regulator $\ell$ and allow the existence of a black hole horizon across a realistic, astrophysical mass range.

In this Note, we show how one may obtain a different outcome by allowing the regulator $\ell$  to have a nontrivial mass dependence.   Put in another way, the quantity $L^3$ in Eq.~(\ref{eq:hayward}) is conventionally assumed to be mass dependent, but the present work explores the phenomenological implications of relaxing the assumption that this dependence is linear.  In doing so, we  obtain regular black hole metrics without the problems related to mass gaps, featuring 
$\mathcal{O}(50\%)$ corrections at the horizon scale in mass ranges associated with astrophysical black holes.   
Given our assumptions about the mass dependence of the regulator, the motivation for considering kilometer-scale values of $\ell$ is similar to the motivation for considering large compactification radii in extra-dimensional scenarios.  When the large extra dimensions idea was first proposed~\cite{Dienes:1998vg,Arkani-Hamed:1998jmv}, there was no theoretical derivation that indicated a priori that one or more compactification radii should be at the millimeter or  the inverse-TeV scale.  The proposal was made to challenge a theoretical prejudice, namely that all compactification scales should be set by the Planck scale.  Dispensing with this preconception led to significant phenomenological benefits, most notably the elimination of the hierarchy problem.  In the present context, there is no theoretical derivation that indicates a priori that the gravitational regulator $\ell$ should be kilometer-scale in size.   Our scenario challenges a theoretical prejudice that the regulator $\ell$ should be set by the Planck scale.  Dispensing with this preconception also leads to significant  phenomenological benefits, most notably the amelioration of the black hole mass gap problem and the possibility of observable horizon-scale effects.  The location of the innermost circular orbit for light (the ``photon sphere''), gravitational lensing, and the black hole shadow, are all susceptible to horizon-scale deviations  in our scenario.  The purpose of this Note is to illustrate these points, which have not been made previously in the literature.  Nevertheless, fully realistic black holes must also address other issues, such as geodesic completeness and stability, and take into account black hole angular momentum.  We discuss these briefly in the final section, as motivation for future work.

\begin{table}[!b]
\centering
\bgroup
\def\arraystretch{1.6}
\begin{tabular}{lclcc} \hline \hline 
Regular black hole model & \hspace{20pt} & Mass gap & \hspace{10pt} & Ref. \\ \hline
Bardeen & & $GM \ge 1.30 \ell $ & & \cite{Bardeen:1968}\\
Dymnikova &  & $GM \ge 0.88 \ell $ & & \cite{Dymnikova:1992ux} \\
Bonanno--Reuter ($\gamma=9/2$) & & $GM \ge 3.50\ell $ & & \cite{Bonanno:2000ep} \\
Hayward & & $GM \ge 1.30 \ell $  & & \cite{Hayward:2005gi} \\
Simpson--Visser & & $GM \ge 0.50 \ell$ & & \cite{Simpson:2018tsi} \\
Frolov ($n=1$) & & $GM \ge 0.98 \ell $ & & \cite{Frolov:2021vbg} \\ \hline \hline
\end{tabular}
\egroup
\caption{Mass gaps for several regular black hole models.}
\label{table:1}
\end{table}

\emph{Modified metric.}---We take the Hayward metric \eqref{eq:hayward} as a starting point, and aim to explore its dependence on the regulator scale $\ell$ as well as its mass parameter $M$ via the dimensionless combination
\begin{align} \label{eq:dimcomb}
\hat{\ell} = \frac{\ell}{2GM} \, .
\end{align}
While $\hat{\ell} \ll 1$ is typically assumed in the study of regular black holes, our focus will be on the regime $\hat{\ell} \gtrsim 1$ where one has strong deviations from the Schwarzschild form away from the origin. We parametrize
\begin{align}
\label{eq:new}
F(r) &= 1 - \frac{2GM}{r} \frac{r^3}{r^3 + L^3} \, , \quad L^3 = 2GM\ell^2 f(\hat{\ell}) \, ,
\end{align}
where the asymptotic ADM mass of this metric is still given by $M$, but $f(\hat{\ell}) > 0$ introduces a new regulator $L$ with an unconventional mass dependence. The black hole horizon, in terms of $\hat{r} \equiv r/(2GM)$, lies at $\hat{r}_\text{h}^3 - \hat{r}_\text{h}^2 = - \hat{\ell}^2 f(\hat{\ell})$. This equation has positive solutions if
\begin{align}
\label{eq:bh-condition}
\hat{\ell}^2 f(\hat{\ell}) \equiv \hat{L}^3 \leq \frac{4}{27} \, ,
\end{align}
which one may think of as the ``black hole condition.'' As in Eq.~(\ref{eq:dimcomb}), and henceforth, a hat indicates that the given quantity has been made 
dimensionless by dividing by an appropriate power of $2 G M$.  The mass-dependence of this condition leads to strikingly different results compared to the Hayward metric. Namely, non-monotonic functions $\hat{\ell}^2 f(\hat{\ell})$ that increase and decrease at intermediate values for $\hat{\ell}$ generate entirely new, astrophysically viable branches for regular black hole metrics that feature both a horizon \emph{and} a large value of $\ell$. To illustrate this, the reader may verify that the following function,
\begin{align}
f_\text{example} = \frac{1}{1+\hat{\ell}^4} \, ,
\label{eq:example}
\end{align}
provides such an example. The black hole condition $\hat{L}^3 \leq 4/27$ is guaranteed for $GM \gtrsim 1.28 \, \ell$,  qualitatively similar to the Hayward case. However, black holes also exist for $GM \lesssim 0.20\,\ell$---an entirely new branch. It is significant that this branch constitutes an \emph{upper} bound on black hole mass,
one that allows black holes with masses that are less than the regulator scale (in units where $c=G=1$). This solves the problem of a disappearing horizon for large regulator scales, so that large effects on black hole observables can be consistently obtained. The choice of function $f(\hat{\ell})$ defines a new family of regular black holes,  as we discuss in some more detail below.

\begin{figure}[!t]
\centering
\includegraphics[width=0.48\textwidth]{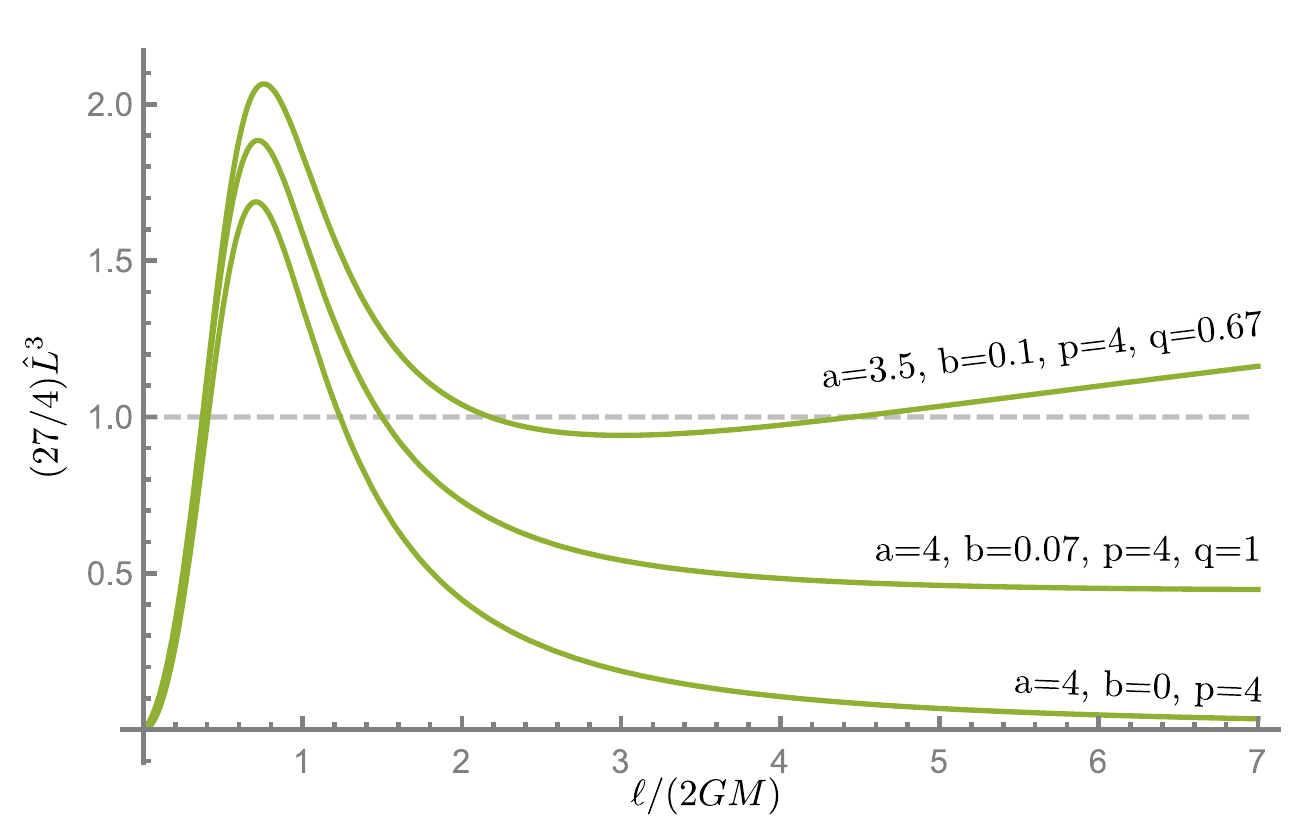}
\caption{Possible parametrizations of the mass-dependent regulator $\hat{L}^3 \equiv \hat{\ell}^2 f(\hat{\ell})$. The black hole condition \eqref{eq:bh-condition} is satisfied whenever the curves are below the dashed line. Given a regulator $\ell$, this implies a band structure for black hole existence.}
\label{fig:parametrization}
\end{figure} 

The impact of this new, mass-dependent regulator can best be captured by estimating horizon-scale effects. Subject to the condition \eqref{eq:bh-condition}, the horizon is located at
\begin{align}
\hat{r}_\text{h} = \frac13\left\{1 + 2\cos\left[\tfrac13\arccos\left(1-\tfrac{27}{2}\hat{L}^3\right)\right]\right\} \, .
\end{align}
At the horizon, the deviation of this class of metrics from the Schwarzschild metric is governed by the ratio
\begin{align}
\delta \equiv \left( \frac{\hat{L}}{\hat{r}_\text{h}}\right)^3 ~ \in ~ [0, 0.5] \, .
\end{align}
The deviation is monotonically increasing and assumes its maximum value of $50\%$ at $\hat{L}^3 = 4/27$. There, usual regular black hole metrics cease to describe black holes and instead become horizonless. The function $f(\hat{\ell})$ in the mass-dependent regulator $\hat{L}$ changes the allowed mass ranges for black holes, so that $\mathcal{O}(50\%)$ horizon-scale effects for astrophysical black holes are allowed;  Fig.~\ref{fig:f} illustrates the effect of the function $f(\hat{\ell})$.  Let us
now constrain its properties a bit more:

\begin{itemize}
\item For a vanishing regulator at fixed mass $M$ ($\hat{\ell}\rightarrow 0$) we should recover the Schwarzschild metric. Assuming that $f(\hat{\ell})$ is regular at the origin, this constrains $f \sim \hat{\ell}^p$ with $p > -2$ at $\hat{\ell}=0$.
\\[-1.4\baselineskip]
\item Close to the origin the metric and its Kretschmann scalar behave as
\begin{align}
\begin{split}
F(r \rightarrow 0) &= 1 - \frac{1}{f} \frac{r^2}{\ell^2} + \mathcal{O}(r^5) \, , \\
K(r \rightarrow 0) &= \frac{1}{f^2} \frac{24}{\ell^4} + \mathcal{O}(r^3) \, ,
\end{split}
\end{align}
where $K \equiv R{}_{\mu\nu\rho\sigma}R{}^{\mu\nu\rho\sigma}$ and we suppressed the argument of $f = f(\hat{\ell})$ for brevity. The limiting curvature condition demands $f$ to be a universal constant to avoid any mass dependence in the maximum curvature. However, if we are willing to allow for a mild mass dependence that still prevents trans-Planckian curvatures, we can relax this condition and merely demand that $f$ approaches a constant as $M$ is taken large with $\ell$ fixed, or equivalently $f(\hat{\ell}) \rightarrow \mbox{constant}$ as $\hat{\ell} \rightarrow 0$.
\item Tabletop experiments for small masses probe the region $M \rightarrow 0$, corresponding to $\hat{\ell} \rightarrow\infty$ at fixed $\ell$. To avoid a more stringent bound than given in Eq.~\eqref{eq:bound}, we assume that $f \lesssim 1$ in this limit.
\end{itemize}
We propose the following, rather general parametrization that captures this essence (see Fig.~\ref{fig:parametrization} for a few cases):
\begin{align}
\label{eq:parametrization}
f(\hat{\ell}) = \frac{1}{1 + a \hat{\ell}^p} + \frac{b}{(1 + \hat{\ell})\hat{\ell}^q} \, .
\end{align}

In the black hole domain, as described by Eq.~\eqref{eq:bh-condition}, the modification is strongest wherever $\hat{L}^3$ reaches its maximum value of $4/27$.   For example, in the case where $a=1$, $p=4$ and $b=0$, \textit{i.e.}, Eq.~\eqref{eq:example}, maximal effects are obtained for black hole masses near the boundaries of the allowed regions, $\sim 0.20 \ell /G$ and $\sim 1.28 \ell / G$, while allowing black holes to exist in both intervals. For $\ell \sim 30\,\text{km}$, these boundaries correspond to $\sim 4 M_\odot$ and $\sim 26 M_\odot$, respectively.    The forbidden mass interval in between can be modified by a different choice of the function $f(\hat{\ell})$. Note that regular supermassive black holes are not affected by the modification to the metric.  Should our qualitative results persist when angular momentum is introduced, the additional stellar black hole ``branches'' with substantial horizon-scale effects suggest new and potentially interesting targets for direct astrophysical observation.

\begin{figure}[!t]
\centering
\includegraphics[width=0.48\textwidth]{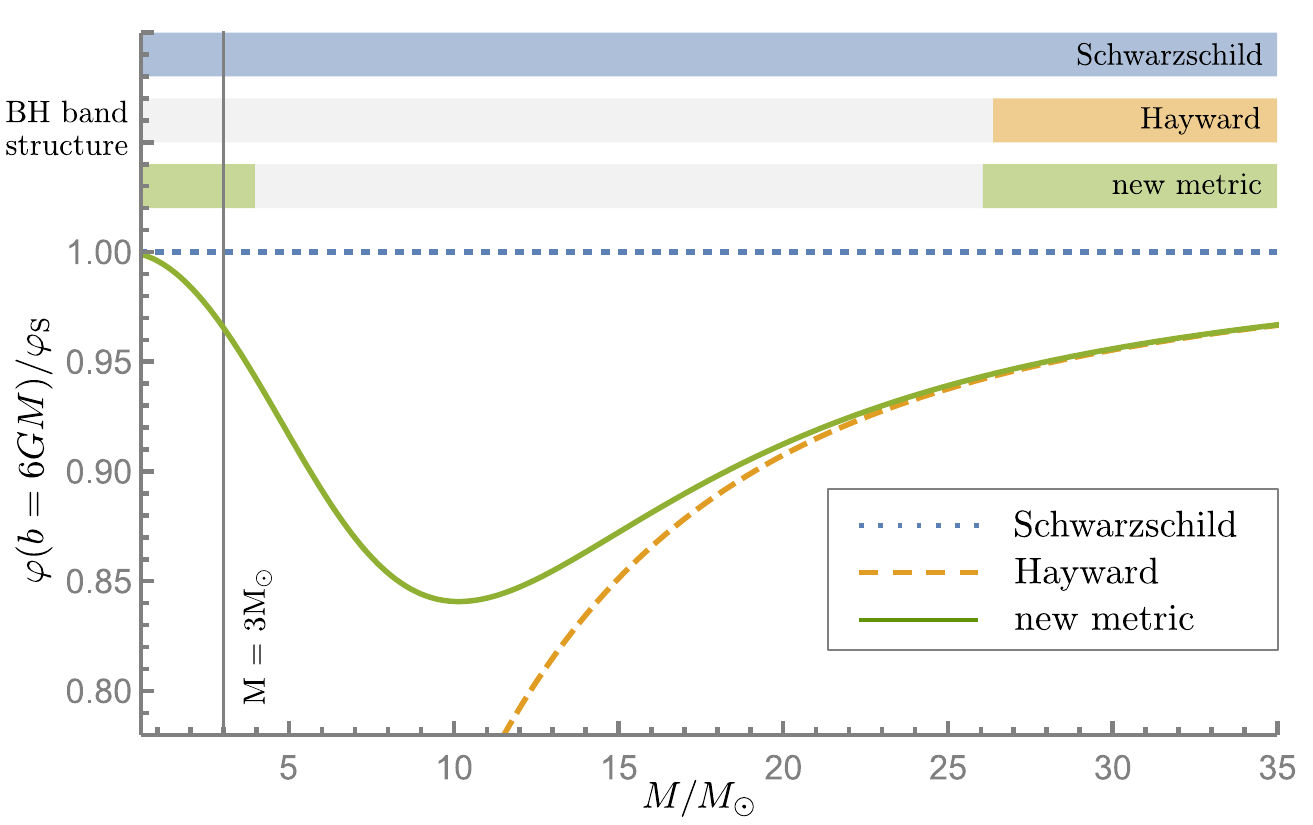}
\caption{Mass-dependent deflection angle $\varphi$ (normalized to the Schwarzschild value) for the impact parameter $b = 6GM$, as a function of mass $M$, assuming Eq.~\eqref{eq:example}. The bands show the allowed black hole mass ranges.}
\label{fig:light-deflection}
\end{figure}

\emph{Consequences.}---To demonstrate the qualitatively new features of the metrics described in this Letter, let us set  \begin{align}
\ell \approx 30\,\text{km}
\end{align}
and explore observational consequences for solar-mass black holes. For simplicity, we will assume a mass-dependent regulator given by $\{a=1, p=4, b=0\}$.

Light propagation in the metric \eqref{eq:new} is described by
\begin{align}
\dot{r}^2 = E^2 - V_\text{eff}(r) \, , \quad V_\text{eff}(r) = \frac{J^2}{r^2}F(r) \, ,
\end{align}
where $E \equiv - g_{tt}\dot{t}$ and $J \equiv g_{\varphi\varphi}\dot{\varphi}$ are constants of motion related to the impact parameter $b$ via $b = J/E$, and the dots denote differentiation with respect to the affine parameter of the geodesic $x{}^\mu(\lambda)$.

The effective potential $V_\text{eff}(r)$ has a maximum outside the black hole horizon at $r = r_\gamma$, indicating an unstable circular photon orbit, called the ``photon sphere.'' The regulator $f(\hat{\ell})$ pushes it inwards, away from $3GM$:
\begin{align}
\label{eq:eom}
\hat{L}^3 = \left(\sqrt{\frac{3GM}{r_\gamma}} - 1\right) \bigg(\frac{r_\gamma}{2GM}\bigg)^3 \ge 0 \, .
\end{align}
The effect is largest for $\hat{L}^3 = 4/27$, resulting in $r_\gamma \approx 2.65\, GM$. This is a 12\% deviation from the Schwarzschild location $3GM$. A similar statement holds for light deflection around such black holes. Following \eqref{eq:eom}, the deflection angle is
\begin{align}
\varphi(b) = 2b \!\!\! \int\limits_{r_0(b)}^\infty \frac{\dd r}{r^2} \frac{1}{\sqrt{1 - \frac{b^2}{r^2} F(r)}} - \pi \, ,
\end{align}
where $r_0$ denotes the point of closest approach, related to the impact parameter $b$ via $F(r_0)\,b^2 = r_0^2$. We can now compare the light deflection around objects of the same mass $M$, described by different metrics. For definiteness, we fix the impact parameter to be twice the location of the Schwarzschild photon sphere. The maximum resulting effect is of $\mathcal{O}(5\%)$, and is displayed in Fig.~\ref{fig:light-deflection}.

We can now define the ``black hole shadow'' as the impact parameter for which the bending of light angle grows to infinity---the shadow is the boundary between trapped and deflected light, and hence corresponds to the visible size of a black hole in the sky. It is given by
\begin{align}
b_\gamma = \frac{r_\gamma}{\sqrt{F(r_\gamma)}} = \frac{r_\gamma}{\sqrt{1-\sqrt{\frac{4GM}{3r_\gamma}}}} \, ,
\end{align}
and the maximum possible deviation from Schwarzschild is again of $\mathcal{O}(5\%)$. In the future, it would be interesting to extend these studies similar to Ref.~\cite{Carballo-Rubio:2018jzw}.

\emph{Conclusions.}---We have described a new class of regular black hole metrics that do not imply a lower bound on the black hole mass below which one 
would obtain a compact horizonless object.  This is made possible by replacing the regulator $\ell$ by a mass-dependent function, entering the metric Eq.~(\ref{eq:hayward}) via
\begin{align}
L^3 = 2GM \ell^2 f\left(\frac{\ell}{2GM}\right) \, ,
\end{align}
where the function $f$ is the new ingredient.   Previously excluded mass ranges are now accessible and permit black hole geometries featuring horizons, while other limited ranges remain inaccessible, without necessarily implying conflict with observation.  The mass dependence allows the regulator $\ell$ and the horizon scale to be {\it comparable}, so that horizon-scale deviations from the black holes described by general relativity at the percent level can now be obtained. These include, but are probably not limited to, a reduction in the black hole's apparent size, a smaller photon sphere, and weaker lensing.

Effects on supermassive black holes are negligible given an appropriate choice of $f$. Since their mass scale exceeds that of astrophysical black holes up to a power of $10^9$~\cite{Bambi:2017iyh}, the effects described in this Letter are suppressed by roughly this factor. This is desirable, since the shadows of supermassive black holes are directly observable \cite{EventHorizonTelescope:2019dse}, and large deviations from general relativity are likely ruled out at that scale. In fact, the observations presently cannot distinguish between black holes from Einstein gravity and modified, regular black hole metrics~\cite{EventHorizonTelescope:2022xqj,Vagnozzi:2022moj}.

The purpose of this Note has been to illustrate the potential usefulness of mass-dependent black hole regulators in the simplest setting, through modification of the form of well-known Hayward metric.  However, a fully realistic application will have to address a number of issues that we reserve for future work:

$\bullet$  {\it Geodesic Completeness and Stability}.
In the recent years, it has been pointed out that several regular black hole models are geodesically incomplete \cite{Bambi:2016wdn,Carballo-Rubio:2019fnb,Zhou:2022yio} or may suffer from instability issues \cite{Carballo-Rubio:2018pmi} related to mass inflation \cite{Poisson:1989zz} at their inner horizons. While our modification does not remedy this behavior, we emphasize that our proposed mass-dependent regulator can be applied to \textit{any} regular black hole model, including models that have improved behavior \cite{Carballo-Rubio:2022kad}.

$\bullet$  {\it Angular Momentum}. 
Realistic astrophysical black holes have angular momentum. The static metric presented in this Note can be endowed with angular momentum following the well-established procedure presented in Ref.~\cite{Azreg-Ainou:2014pra}; alternatively, one may replace the regulator in known rotating regular black hole metrics \cite{Simpson:2021dyo} directly with the mass-dependent regulator proposed in this Note.

$\bullet$ \textit{Energy Conditions.} The absence of singularities in regular black hole models is typically accompanied with the violation of energy conditions, which has been studied in great detail elsewhere \cite{Zaslavskii:2010qz,Balart:2014jia}. In the context of the proposed model---where the radial dependence of the metric coincides with that of the Hayward metric---the only difference emerges through the re-scaled regulator:
\begin{align}
L = (2GM)^{1/3} \ell^{2/3} f^{1/3} \, .
\end{align}
Recall that for the astrophysical mass range, kilometer-scale fundamental regulators $\ell$ would give an $L$ that is also of the order of kilometers---were it not for the modification function $f$ that can be chosen to reduce the total regulator $L$. Since the scale $L$ sets the violation length scale for the energy conditions, in the proposed model the violation will be constrained to that modified range, which, for astrophysically relevant black holes, is smaller than in the pure Hayward case. In this case, this model therefore shrinks the effective regions of energy condition violations. However, we also note that energy conditions, in their standard form, only apply to general relativity, wherein the energy-momentum tensor is algebraically related to the Ricci curvature. In modified gravity theories (or, for example, an effective theory of quantum gravity that may dynamically generate regular black hole metrics like the one discussed in this Note) this algebraic relation may no longer be true. In the absence of an action principle and equations of motion, such questions cannot be unambiguously answered.\footnote{This also applies to stability issues related to black hole thermodynamics, since the latter as a semiclassical approach can only be defined in the presence of an action.}

$\bullet$ {\it Gravitational Waves}. 
Since the discussion in this Letter is limited to time-independent metrics, one may wonder about the dynamical aspects of these black holes and their non-trivial band structures ({\it i.e.}, allowed mass intervals). For example, is it possible to have two black holes from the low-mass part of the spectrum, make them collide, and end up in the regime where no black holes are allowed? And, in that case, what is the nature of the horizonless end product after a collision of these black holes? It is well known that such compact, horizonless object can feature instabilities \cite{Cunha:2022gde}. This may then lead to emission of gravitational waves that only terminates once the object is again inside the lower mass range  and a horizon has formed. In the presence of more involved regulators $f(\hat{\ell})$ one could have multiple, disconnected bands for these types of black holes, giving rise to a multitude of transitions between bands, accompanied by a characteristic amount of gravitational wave emission. While certainly interesting, we will leave dynamical aspects to future work.

Lastly, we point out that both the unmodified Hayward metric and the metrics proposed in this Note satisfy the limiting curvature condition \cite{Markov:1982,Markov:1984,Polchinski:1989}, which is rooted in quantum-gravitational considerations. Thus, one might hope that regulators with an unconventional mass dependence
may one day be traced back to an underlying theory of quantum gravity.\footnote{One might view the unconventional mass dependence of the regulator in Ref.~\cite{Bonanno:2000ep} as
a step in this direction.}

\textit{Acknowledgements.}---We are grateful for support by the National Science Foundation under grant no.~PHY-2112460. J.B. acknowledges support as a Fellow of the Young Investigator Group Preparation Program administered by the state of Baden--W\"urttemberg (Germany) and the Karlsruhe Institute of Technology.

\appendix

\end{document}